\documentstyle[12pt]{article}
\begin{document}
\setlength{\unitlength}{1mm}
\newcommand{\te}{\theta}
\renewcommand{\thefootnote}{\fnsymbol{footnote}}
\newcommand{\bee}{\begin{equation}}
\newcommand{\ene}{\end{equation}}
\newcommand{\bea}{\begin{eqnarray}}
\newcommand{\ena}{\end{eqnarray}}
\newcommand{\dx}{d(x)}
\newcommand{\ux}{u(x)}
\newcommand{\sx}{s(x)}
\newcommand{\tra}{\triangle\theta_}
\newcommand{\uupv}{u^{\uparrow}_{val}}
\newcommand{\udwv}{u^{\downarrow}_{val}}
\newcommand{\dupv}{d^{\uparrow}_{val}}
\newcommand{\ddwv}{d^{\downarrow}_{val}}
\newcommand{\auupv}{\bar{u}^{\uparrow}_{val}}
\newcommand{\audwv}{\bar{u}^{\downarrow}_{val}}
\newcommand{\adupv}{\bar{d}^{\uparrow}_{val}}
\newcommand{\addwv}{\bar{d}^{\downarrow}_{val}}
\newcommand{\uupx}{u^{\uparrow}(x)}
\newcommand{\udwx}{u^{\downarrow}(x)}
\newcommand{\dupx}{d^{\uparrow}(x)}
\newcommand{\ddwx}{d^{\downarrow}(x)}
\newcommand{\auupx}{\bar{u}^{\uparrow}(x)}
\newcommand{\audwx}{\bar{u}^{\downarrow}(x)}
\newcommand{\adupx}{\bar{d}^{\uparrow}(x)}
\newcommand{\addwx}{\bar{d}^{\downarrow}(x)}
\newcommand{\fnx}{{F_{2}^{n}}(x)}
\newcommand{\fpx}{{F_{2}^{p}}(x)}
\newcommand{\aux}{\bar{u}(x)}
\newcommand{\adx}{\bar{d}(x)}
\newcommand{\asx}{\bar{s}(x)}
\newcommand{\gpx}{{g_{1}^{p}}(x)}
\newcommand{\gnx}{{g_{1}^{n}}(x)}

{{\hfill \small Universit\'a di Napoli Preprint DSF-T-43/94 (december 1994)}}

\begin{center}
{\Large\bf Fermi-Dirac statistics plus liquid description of quark partons}
\end{center}
\bigskip\bigskip

\begin{center}
{{\bf F. Buccella}$^{1}$, {\bf G. Miele}$^{1,2}$, {\bf G. Migliore}$^{1}$
and {\bf V. Tibullo}$^{1}$}
\end{center}

\bigskip

\noindent
{\it $^{1}$ Dipartimento di Scienze Fisiche, Universit\`a di Napoli
''Federico II'', Mostra D'Oltremare Pad. 20, I--80125 Napoli, Italy}\\

\noindent
{\it $^{2}$ Istituto Nazionale di Fisica Nucleare, Sezione di Napoli,
Mostra D'Oltremare Pad. 20, I--80125 Napoli, Italy}\\

\bigskip\bigskip\bigskip

\begin{abstract}
A previous approach with Fermi-Dirac distributions for fermion partons is
here improved to comply with the expected low $x$ behaviour of structure
functions. We are so able to get a fair description of the unpolarized
and polarized structure functions of the nucleons as well as of neutrino
data. We cannot reach definite conclusions, but confirm our suspicion of a
relationship between the defects in Gottfried and spin sum rules.
\end{abstract}

\vspace{5cm}

\centerline{To appear in {\it Zeit. f\"{u}r Phys. C}}

\baselineskip22pt
\newpage

\section{Introduction}
The experimental data on the unpolarized and polarized structure
functions of the nucleons suggest a role of Pauli principle in
relating the shapes and the first momenta of the distributions of the
various quark parton species, for which Fermi-Dirac distributions in
the $x$ variable have been proposed. Here, to comply with the expected
behaviour for the structure functions in the limit $x
\rightarrow 0$, we add a {\it liquid} unpolarized component dominating the
very low $x$ region and not contributing to quark parton model sum rules
(QPMSR).
Section 2 deals with the motivations for our description of parton
distributions and the comparison with available data. The results
found and the implications for QPMSR are discussed in section 3. Finally
we give our conclusions.

\section{Pauli exclusion principle and the parton distributions}

The behaviour at high $x$ of $\fnx/\fpx$ \cite{1}, known since a long time,
and the more recent polarization experiments \cite{2}, \cite{2bis},
which show that at high $x$ the partons have spin parallel to the one
of the proton, imply that $\uupx$ is the dominating parton distribution in
the proton at high $x$.\\
Indeed at $Q^2 =0$, the axial couplings of the baryon octet are fairly
described in terms of the valence quarks
\bee
\uupv  = 1+F~~~~~~~~~~~~\udwv  = 1-F~~~,
\label{1}
\ene
\bee
\dupv  = {1+F-D \over 2}~~~~~~~~~~\ddwv  = {1-F+D \over 2}~~~.
\label{2}
\ene
With the actual values \cite{pdg94}, \cite{3}
\bee
F=0.464 \pm 0.009~~~~~~~~~~~D=0.793 \pm 0.009~~~,
\label{3}
\ene
$\uupv$ is larger that the others
\bee
\uupv \approx { 3 \over 2} \approx \udwv + \dupv + \ddwv~~~.
\label{4}
\ene
In a previous work we assumed that the parton distributions at a
given large $Q^2$ depend on their abundance at $Q^2 =0$ \cite{4}
\bee
p(x) = {\cal F}(x,p_{val})~~~,
\label{5}
\ene
with ${\cal F}$ an increasing function of $p_{val}$ and with a broader shape
for higher values of $p_{val}$.\\ This assumption and the observation
that with $F=1/2$ and $D=3/4$  (very near to the values quoted in
Eq.(\ref{3})) one has $\udwv = 1/2$ just on the center of the narrow
range $(\dupv, \ddwv) \equiv (3/8,5/8)$ lead to take \cite{5}
\bee
\udwx = { 1 \over 2} \left[ \dupx + \ddwx \right] = { 1 \over 2}
\dx~~~,
\label{6}
\ene
which implies
\bee
\Delta\ux = \uupx - \udwx = \ux - \dx~~~.
\label{7}
\ene
Eq.(\ref{7}) connects the contribution of $\Delta\ux$ to the
polarized structure function of the proton $\gpx$, with the
contributions of $u$ and $d$ to the difference between the proton and
the neutron unpolarized structure functions $F_{2}(x)$ \cite{5}
\bee
x \gpx \Bigr|_{\Delta u} = {2 \over 3} \left[ \fpx -\fnx
\right]_{u+d}~~~.
\label{8}
\ene
Since a smaller negative contribution to $\gpx$ is expected from
$\Delta\dx$ for the twofold reason that $e_{d}^2 = (1/4) e_{u}^2$ and
$\Delta d_{val} \approx - (1/4) \Delta u_{val}$, Eq.(\ref{8})
should hold with a good approximation for the total structure
functions in the region dominated by the valence quarks: this is just
the case for $x \geq 0.2$ \cite{5}.

By integrating Eq.(\ref{7}) one relates $\Delta u$, which contributes
to the spin sum rules, to $u-d$ contributing to the Gottfried sum rule
\cite{6}
\bee
I_{G} = \int_{0}^{1} { 1\over x}\left[\fpx - \fnx \right]~dx = {
1\over 3} ( u + \bar{u} -d -\bar{d} ) = { 1 \over 3}~~~,
\label{9}
\ene
where the last equality follows if the sea is $SU(2)_{I}$ invariant
($\bar{d} = \bar{u}$). Indeed NMC experiment \cite{7} gives for the
l.h.s. of Eq.(\ref{9})
\bee
I_{G} = 0.235 \pm 0.026~~~,
\label{10}
\ene
implying
\bee
\bar{d} - \bar{u} = 0.15 \pm 0.04 ~~~~~~~~~~~~ u - d = 0.85 \pm 0.04~~~.
\label{11}
\ene
Many years ago Field and Feynman suggested \cite{8} that Pauli
principle disfavours the production of $u \bar{u}$ pairs in the proton
with respect to $d \bar{d}$, since it contains two valence $u$ quarks
and only one $d$. The correlation shape-abundance for the parton
distributions is also the one suggested by the Pauli principle: the
most abundant parton $u^{\uparrow}$ is the one dominating at high $x$
and the assumption that $u^{\downarrow}$ and $d$ have about the same
shape seems confirmed by the experiment.

The role of the Pauli principle has suggested to assume
Fermi-Dirac distributions in the variable $x$ for the quark partons \cite{4}
\bee
p(x) = f(x) \left[\exp\left({ x - \tilde{x}(p) \over \bar{x}} \right) + 1
\right]^{-1}~~~,
\label{12}
\ene
where $f(x)$ is a weight function, $\bar{x}$ plays the role of the temperature
and $\tilde{x}(p)$ is the {\it thermodynamical potential} of the parton
$p$, identified  by its flavour and spin direction.\\
Consistently for the gluons, neglecting their polarization, one
assumed \cite{4}
\bee
G(x) = {16 \over 3} f(x)
\left[\exp\left({ x - \tilde{x}_{G} \over \bar{x}} \right) - 1
\right]^{-1}~~~,
\label{13}
\ene
where the factor $16/3$ is just the product of $2$ ($S_{z}(G) = \pm 1$)
times $8/3$, the ratio of the colour degeneracies for gluons and
quarks. To reduce the number of parameters, the distributions of $d$, $s$ and
of their antiparticles have been given in terms of the ones for $u$ and
$\bar{u}$ \cite{4}
\bea
\dx & = & {\uupx \over 1-F}~~~,\label{14}\\
\Delta\dx & = & - k~f(x)~\exp\left({ x - \tilde{x}(u^{\downarrow})
\over \bar{x}}\right)
\left[\exp\left({ x - \tilde{x}(u^{\downarrow}) \over \bar{x}} \right) + 1
\right]^{-2}~~~,\label{15}\\
\adupx & = & \addwx = \audwx~~~, \label{16}\\
s(x) & = & \bar{s}(x) = { 1 \over 4} \left[ \aux + \adx \right]~~~,\label{17}\\
\Delta s(x) & = & \Delta \bar{s}(x) =0~~~,\label{18}
\ena
with $k$ fixed by the condition $\Delta d=\Delta d_{val} =F-D$.
For the weight
function one considered the simple form $f(x) = A~x^{\alpha}$. In terms of
seven parameters, $\bar{x}$, the $\tilde{x}$ for $u$ and $\bar{u}$, $A$ and
$\alpha$, one obtains a nice description of the unpolarized and polarized
structure functions for the nucleons, but it is not possible to reproduce the
the fast increasing \cite{9} at low $x$ of $\bar{q}(x)=\aux+\adx+\bar{s}(x)$,
confirmed at very high $Q^2$ from the behaviour of $\fpx$ measured at $H_{1}$
\cite{10}.

Indeed, we know that the form given in Eq.(\ref{12}), with different values
of the $\tilde{x}$ for the different parton species, is not suitable
in the limit $x \rightarrow 0$ for the most divergent part expected
on general grounds to be equal for the different partons, at least in
the limit of flavour symmetry. This most divergent part should not
contribute to QPMSR as the ones given by Gottfried and Bjorken
\cite{11} with $I=1$ quantum numbers exchanged. It is therefore needed,
to reproduce the data and to get information on the status of QPMSR
within this approach, to disentangle the most divergent part for $x
\rightarrow 0$
of the parton distributions (which does not contribute to QPMSR) from
the remaining part, just given by Eq.(\ref{12}).

To this extent we add a {\it liquid} unpolarized
component, giving to the light partons,
$u$, $d$ and their antiparticles, the same contribution $f_{L}(x) = A_{L}~
x^{\alpha_{L}} (1-x)^{\beta_L}$, and to $s$ and $\bar{s}$, as in Eq.(\ref{17}),
$f_{L}(x)/2$. To be not influenced by theoretical prejudices we
consider as free parameters the $\tilde{x}$ for $u^{\uparrow(\downarrow)}$,
$d^{\uparrow(\downarrow)}$, $\bar{u}$ and
$\bar{d}$. Finally we introduce a new parameter in $f(x)$
\bee
f(x)= A~ x^{\alpha} (1 -x)^{\beta}~~~.
\label{19}
\ene
We try also to describe the structure function
\bee
F_{3}(x) = \ux+\dx+\sx-\aux-\adx-\asx~~~,
\label{20}
\ene
measured with great precision in deep inelastic reactions induced by
(anti)neutrinos \cite{12}.\\
According to Eq.(\ref{17}), we expect $\sx -\asx=0$ and thus $\fpx -\fnx$ and
$F_{3}(x)$ to depend on the differences $\ux-\adx$ and $\dx - \aux$. Note that
we cannot impose the conditions
\bea
u - \bar{u} & = & 2~~~,\label{21}\\
d - \bar{d} & = & 1~~~,\label{22}
\ena
since they would imply the Gross and Llewellyn-Smith sum rule \cite{13}
\bee
\int_{0}^{1} F_{3}(x)~dx=3~~~,
\label{23}
\ene
which as well-known experimentally shows a defect. The l.h.s. of
Eq.(\ref{23}) is in fact measured to be
$2.50 \pm 0.018~({\mbox{stat.}}) \pm 0.078~({\mbox{syst.}})$ \cite{12},
defect commonly explained in terms of QCD corrections \cite{17}.\\

In Figures 1-6 we compare our predictions for
$\fpx -\fnx$, $xF_{3}(x)$, $x\gpx$ and
$x\gnx$, which do not receive contributions from the {\it liquid}
component, and for $\fnx/\fpx$ and $x\bar{q}(x)$ with the experiments.
We restrict the $\tilde{x}$'s to be $\leq 1$, since the factor $(1-x)^{\beta}$
in $f(x)$ makes the dependence of the distributions on $\tilde{x}\geq
1$ very smooth.

Our initial goal was to introduce spin-dependent $\tilde{x}$'s
also for the $\bar{q}$'s and to test the relationship
\bee
\Delta \bar{u}(x) = \bar{u}(x) - \bar{d}(x)~~~,
\label{1a}
\ene
assumed in previous works (\cite{5} and \cite{14}); Eqs.(\ref{7}) and
(\ref{1a}) would imply the equality
\bee
x \gpx \Bigr|_{\Delta u+\Delta \bar{u}} = {2 \over 3} \left[ \fpx -\fnx
\right]~~~.
\label{2a}
\ene
Unfortunately, we found practically the same $\chi^2$ with negative
and positive values for $\Delta\bar{u}(x)$ and/or $\Delta\bar{d}(x)$,
and realized that, with $f(x)$ to be found from the data, we are unable
to disentangle the contributions of $\Delta q(x)$ and $\Delta \bar{q}(x)$
to the polarized structure functions. Thus, our choice $\Delta \bar{u}(x)=
\Delta \bar{d}(x)=0$ is neither motivated by data, nor by theoretical
prejudices, but simply from our present inability to get information even on
their signs and to settle the important issue, relevant also for the validity
of the Bjorken sum rule, whether Eq.(\ref{1a}) is satisfied. Eq.(\ref{2a}),
with only $\Delta u$, $\Delta d$, $\Delta \bar{u}$ and $\Delta \bar{d}$
contributing to the polarized structure functions of the nucleons, would
imply
\bee
x \left[ \gpx - { 1 \over 4} \gnx \right] = {5 \over 8} \left[ \fpx -\fnx
\right]~~~.
\label{3a}
\ene
In Table 1 we report the parameters found here by means of the MINUIT
fitting code, as well as the
ones of the previous fit (without liquid) and of a one by Bourrely and
Soffer \cite{14} found on similar principles, but with several different
assumptions from ours.

Indeed, the data on unpolarized nucleons structure functions are at
$Q^2 = 4~GeV^2$ \cite{7}, the neutrino data at $Q^2 = 3~GeV^2$ \cite{12},
and $\bar{q}$ measures are performed at $Q^2 = 3~GeV^2$ and $5~GeV^2$
\cite{9} and differ at small $x$, while our curve is intermediate
between the two sets of
data. The data on $g_1^n(x)$ are at $Q^2 = 2~GeV^2$ \cite{16},
whereas $g_1^p(x)$ is measured at $Q^2 = 10~GeV^2$ by SMC \cite{2} and
at $Q^2 = 3~GeV^2$ by E143 \cite{2bis}; despite some narrowing of
the distribution at higher $Q^2$ showing up in the data, the values of
$I_p$ are in good agreement. This fact and the expected $Q^2$
dependance \cite{14}, smaller than the actual errors on the polarized
structure functions, gives us confidence that our analysis is slightly
affected by our neglecting the $Q^2$ dependance of the distributions.

The parton distributions so found are described in Figure 7. The total
momentum carried by $q$ and $\bar{q}$ is $53\%$. In order that gluons
carry out the remaining part of the momentum $\tilde{x}_G$ is fixed to
be $-1/15$. The gluon distribution is compared with the information
found on them in CDHSW \cite{CDHSW}, SLAC$+$BCDMS \cite{SLAC&BCDMS} and
in NMC \cite{NMCgluon} experiments at $Q^2 = 20~GeV^2$ in Figure 8. The
agreement is fair for $x > .1$, while the fast increase at small $x$,
confirmed also from the data at very small $x$ at Hera \cite{H1gluon},
confirms that a liquid component is needed also for gluons. The excess
at high $x$ of our curve with respect to experiment may be, at least
in part, explained by the expected narrowing of the distribution from
$Q^2 = 4~GeV^2$, where we fit the unpolarized distributions,
to $Q^2 = 20~GeV^2$\footnote{Indeed the gluon distributions are
obtained from the $Q^2$ dependance of the distributions according to
the LAP equations \cite{LAP}. In this respect it is worth noticing
that the parameter $\Lambda_{QCD}$ found from the corrections to the
scaling is slightly smaller than the one found from different sources
\cite{pdg94}. This qualitatively supports the idea that the evolution
equations may be modified as a consequence of quantum statistical
effects \cite{APmod}, which would favour harder quarks and softer
gluons, giving rise to a slower softening of quark distributions with
increasing $Q^2$}.

\section{Discussion of the results}

The inclusion of the {\it liquid} term and the extension of our fit to
the precise experimental results on neutrinos has brought to substantial
changes in the parameters with respect to the previous work \cite{4}.

The low $x$ behaviour of $f(x)$ become smoother ($\simeq x^{-.203 \pm .013
}$ instead of $x^{-0.85}$), but this is easily understood since the
previous behaviour was a compromise between the smooth {\it gas}
component and the rapidly changing {\it liquid} one to reproduce the
behaviour of $\bar{q}(x)$. The {\it liquid} component, relevant only
at small $x$, carries only $.6\%$ of parton momentum and its behaviour
$\sim x^{-1.19}$, similar to the result  found in \cite{capella},
is less singular than the one, suggested in the
framework of the multipherial approach to deep-inelastic scattering,
proportional to $\sim x^{-1.5}$ \cite{15}. The parameter
$\tilde{x}(u^{\uparrow})$ took the highest value allowed by us (1.),
since the factor in $f(x)$,
$(1-x)^{2.34}$, is taking care to decrease $\uupx$ at high $x$. The
temperature $\bar{x}$ is  larger than
the previous one and the one found by Bourrely and Soffer \cite{14}. Instead
$\tilde{x}(u^{\downarrow})$ is slightly smaller than the previous
determination \cite{4} and about half the value found in \cite{14},
where $f(x)$ is different for $u^{\uparrow}$ and $u^{\downarrow}$.

The ratio $r = \udwx/\dx$ varies in the
narrow range $(.546,.564)$ in fair agreement with the constant value
$1 -F = .536 \pm .009$ assumed in \cite{4} and
slightly larger than the value $1/2$ taken in \cite{5} and \cite{14}.

The central value found for the first moment of $\bar{u}_{gas}(x)$, $.03$,
is smaller than $\bar{d}_{gas}(x)/2$, $.08$, while Eq.(\ref{1a})
implies $\bar{u}(x) \geq \bar{d}(x)/2$. However, the large upper error on
$\bar{u}_{gas}$ and the uncertainty
in disentangling the gas and liquid contributions for the $\bar{q}$'s do not
allow to reach a definite conclusion about the validity of Eq.(\ref{1a}).\\
Indeed our distributions are very well consistent with Eq.(\ref{3a}), as it is
shown in Figure 9, where our predictions for the two sides of Eq.(\ref{3a})
are compared.

We have been suggested by Prof. Jacques Soffer to compare the parton
distributions found here with the measured asymmetry for Drell-Yan
production of muons at $y=0$ in $pp$ and $pn$ reactions
\bee
A_{DY} = { d \sigma_{pp}/dy - d \sigma_{pn}/dy \over
d \sigma_{pp}/dy + d \sigma_{pn}/dy}~~~,
\label{ady}
\ene
which at rapidity $y=0$ reads
\bee
A_{DY} = { (\lambda_s(x) -1) (4 \lambda(x) -1) +
(\lambda(x) -1) (4 \lambda_s(x) -1) \over
(\lambda_s(x) +1) (4 \lambda(x) +1) +
(\lambda(x) +1) (4 \lambda_s(x) +1) }~~~,
\label{ady1}
\ene
where $\lambda_s(x)=\bar{u}(x)/\bar{d}(x)$ and
$\lambda(x)=u(x)/d(x)$. At $x=.18$ we have $\lambda_s(.18)=.454$
and $\lambda(.18)=1.748$ giving rise to $A_{DY}(.18)=-.138$ in
fair agreement with the experimental result $-.09 \pm .02 \pm .025$
\cite{NA51}.\\
The behaviour of $A_{DY}(x)$ is plotted in Figure 10 together with
the experimental point measured by NA51 collaboration.

We consider now the implications for the QPMSR and we begin with the
one by Gross and Llewellyn-Smith \cite{13}
\bee
\int_{0}^{1} F_{3}(x)~dx = u + d -\bar{u} - \bar{d}=3~~~.
\label{30}
\ene
{}From Table 1 we get for the l.h.s. of Eq.(\ref{30}) $2.44
\begin{array}{c} +.05 \\-.08 \end{array}$
in good agreement with the experimental value $2.50 \pm 0.018~({\mbox{stat.}})
\pm 0.078~({\mbox{syst.}})$.

For the l.h.s. of Gottfried sum rule \cite{6} we get
\bee
{ 1 \over 3} (u + \bar{u} - d - \bar{d}) = .20 \pm .02~~~,
\label{31}
\ene
to be compared with $.235 \pm .026$ \cite{7}.
As long as for the spin sum rules
we get
\bea
\Delta u & = & .62 \pm .02~~~,\nonumber\\
\Delta d & = & -.29 \pm .04~~~,
\label{32}
\ena
to be compared with
\bea
\Delta u_{val} & = & 2F = .93 \pm .02~~~,\nonumber\\
\Delta d_{val} & = & F-D =-.33 \pm .02~~~,
\label{32a}
\ena
{}From Eq.(\ref{32}) we get
\bee
I_{p} =  { 2 \over 9} \Delta u + { 1 \over 18} \Delta d = .122 \pm .007
{}~~~,
\label{33a}
\ene
\bee
I_{n} = { 1 \over 18} \Delta u  + { 2 \over 9}
\Delta d = -.030 \pm .010~~~,
\label{33b}
\ene
consistent with the SMC result $.136 \pm .011 \pm .011$ \cite{2} and
the E143 result $.129 \pm .004 \pm .009$ \cite{2bis} for $I_p$,
and with the E142 result $-.022 \pm .011$ \cite{16} for $I_n$.

For the Bjorken sum rule \cite{11} we get
\bee
I_{p} - I_{n} = { 1 \over 6} (\Delta u - \Delta d) = .152 \pm .010~~~,
\label{34}
\ene
smaller than $(1/6)|g_{A}/g_{V}|=.209$.

In Table 2. we compare our evaluations of l.h.s. QPMSR with the experiment
and the prediction of theory without the QCD corrections.

By comparing the value found for the first momenta (often called by
us more prosastically abundances) of the gas components of the different
parton species with the r.h.s.'s of Eqs.(\ref{1}) and (\ref{2}), one finds
\bea
u^{\uparrow}_{gas} & = 1.15 \pm .01
& < u^{\uparrow}_{val} = 1+F = 1.464 \pm .009~~~,\nonumber\\
d^{\downarrow}_{gas} & = .62 \pm .01
& \leq d^{\downarrow}_{val} = {1+D-F \over 2} = .665 \pm .009~~~,\nonumber\\
u^{\downarrow}_{gas} & = .53 \pm .01
& \approx u^{\downarrow}_{val} = 1-F = .536 \pm .009~~,\nonumber\\
d^{\uparrow}_{gas} & = .33 \pm .03
& \approx d^{\uparrow}_{val} = {1+F-D \over 2} = .335 \pm .009~~~.
\label{35}
\ena
The different behaviour of $u^{\uparrow}$ with respect to the other
valence quark, for which $q_{gas} \sim q_{val}$,
may be understood in the framework described here as an effect
of Pauli blocking, since its levels are almost completely occupied
differently from the other valence quarks with smaller {\it potentials},
as it is also shown by the fact that $\tilde{x}(u^{\uparrow})$
takes the highest value allowed. Thus, the interpretation of the defect in
Gottfried sum rule as a consequence of Pauli principle, disfavouring
the most abundant valence parton, $u^{\uparrow}$, seems supported by
the inequalities (\ref{35}). This interpretation would bring to the
very relevant consequence of a defect in the Bjorken sum rule.
This conclusion is also supported by the
good agreement with the data of the relationship
\bee
u^{\downarrow} = {d \over 2} + { 1 \over 2} -F~~~,
\label{36}
\ene
which implies
\bee
\Delta u = u - d + 2F -1~~~.
\label{36a}
\ene
With the abundances found by us Eq.(\ref{36}) reads
\bee
.53 \pm .01 = .51 \pm .03~~~.
\label{36b}
\ene

A word of caution is welcome for our conclusions on the violation of
Bj sum rule, since we did not include the effect of QCD corrections
in relating the quark parton distributions to the structure functions.
Also we assumed no polarization for $\bar{q}$, being unable to get a
reliable evaluation of $\Delta \bar{q}$ with the present precision
for the polarized structure functions at
small $x$. Indeed our description of $g_1^p(x)$ and $g_1^n(x)$
is good in terms of $\Delta u(x)$ and $\Delta d(x)$, but our
prediction is smaller than the central values of the three lowest $x$
values measured by SMC.

\section{Conclusions}

We compared with data the quark-parton distributions given by the sum of
Fermi-Dirac functions and of a term not contributing to the QPMSR relevant at
small $x$. We obtain a fair description for the unpolarized and polarized
structure functions of the nucleons as well as for the $F_3(x)$ precisely
measured in (anti)neutrino induced deep-inelastic reactions and for $\bar{q}$
total distribution. The conjectures of previous works on $d$ distributions are
well confirmed by the values chosen for their thermodynamical potentials. As
long as the implications for QPMSR the values found for the first momenta of
the various parton species give l.h.s.'s consistent with experiment. For the
fundamental issue of the Bjorken sum rule, as advocated in previous works
\cite{4}, \cite{5} and \cite{19}, we get
\bea
\Delta u & \approx  & u - d + 2F-1~~~,\label{37}\\
\Delta d & \geq  & F- D~~~,\label{38}
\ena
to confirm the suspicion
of a violation of Bjorken sum rule related to
the defect in the Gottfried sum rule.

\newpage
\bigskip\bigskip
\par\noindent
{\bf Table 1.}\\
\bigskip
\begin{tabular}{|c|c|c|c|c|}
\hline
Parameters
& Previous fit  \cite{4} & Fit BS \cite{14}&\multicolumn{2}{c|}{Actual fit}\\
& & & \multicolumn{2}{c|}{$\chi^2/N=2.47$}\\
\hline
$A$ & $.58$ &  & \multicolumn{2}{c|}{$ 2.66\begin{array}{c} +.09\\ -.08
\end{array} $}\\
$\alpha$ & $-.85$ & $\begin{array}{c} -.646~\mbox{for}~u^{\uparrow}_{val}\\
-.262~\mbox{for}~u^{\downarrow}_{val}\end{array} $ &
\multicolumn{2}{c|}{$ -.203 \pm .013$}\\
$\beta$ &  &  &  \multicolumn{2}{c|}{$ 2.34
\begin{array}{c} +.05\\ -.06 \end{array}$}\\
$A_{L}$ &  &  & \multicolumn{2}{c|}{$ .0895 \begin{array}{c} +.0107\\
-.0084\end{array}$}\\
$\alpha_{L}$ & & & \multicolumn{2}{c|}{$ -1.19 \pm .02$}\\
$\beta_{L}$ & & & \multicolumn{2}{c|}{$ 7.66 \begin{array}{c} +1.82\\
-1.59\end{array}$}\\
\cline{4-5}
$\bar{x}$& $.132$ & $.092$ & $.235 \pm .009$ & gas abund. \\
\cline{5-5}
$\tilde{x}(u^{\uparrow})$& $.524$&  $.510~\mbox{for}~u^{\uparrow}_{val}$
& $1.00 \pm .07$ & $1.15 \pm .01$\\
$\tilde{x}(u^{\downarrow})$& $.143$ & $.231
{}~\mbox{for}~u^{\downarrow}_{val}$ & $.123 \pm .012$ &
$.53 \pm .01$\\
$\tilde{x}(d^{\uparrow})$&  & & $-.068 \begin{array}{c} +.021\\
-.024\end{array}$ & $.33 \pm .03$\\
$\tilde{x}(d^{\downarrow})$&  &  &$.200 \begin{array}{c} +.013\\
-.014\end{array}$ & $.62 \pm .01$\\
$\tilde{x}(\bar{u}^{\uparrow})
$& $-.216$ &  & $-.886 \pm .266$ & $.015 \begin{array}{c} +.034\\
-.009\end{array}$\\
$\tilde{x}(\bar{u}^{\downarrow})$& $-.141$ &  & $''$ & $''$\\
$\tilde{x}(\bar{d}^{\uparrow})=\tilde{x}(\bar{d}^{\downarrow})
$& $''$ &  & $-.460\begin{array}{c} +.047\\
-.064\end{array}$ &$.08 \begin{array}{c} +.03\\
-.02\end{array}$\\
\hline
\end{tabular}
\newpage
\bigskip\bigskip
\par\noindent
{\bf Table 2.}\\
\bigskip
\begin{tabular}{|c|c|c|c|}
\hline
Sum rule & Experimental data & Our fit & QPM \\
\hline
GLS & $2.50 \pm .018 \pm .078$ \cite{12}
& $2.44\begin{array}{c}+.04\\-.07\end{array}$ & $3$ \\
G & $.235 \pm .026$ \cite{7}
&$.20 \pm .02$ &$1/3$ \\
EJ $\left\{\begin{array}{c} I_{p}\\ ~ \\ ~ \\ I_{n} \end{array}\right.$ &
$\begin{array}{c} .136 \pm .011 \pm .011~ \cite{2}
\\ .129 \pm .004 \pm .009~ \cite{2bis}
\\ \\ -.022 \pm .011~ \cite{16} \end{array}$ &
$\begin{array}{c} {.122 \pm .007}\\
{-.030 \pm .010}\end{array}$ &
$\begin{array}{c} .188 \pm .005\\ ~ \\-.021 \pm .005\end{array}$ \\
Bj & & $.152 \pm .010$ & $.209$\\
\hline
\end{tabular}
\newpage
{{\Large \bf Table captions}}

\bigskip

\begin{itemize}
\item[Table 1.] Comparison of the values for the parameters of
our best fit with the corresponding quantities, if any, found in
previous analysis \cite{4}, \cite{14}.
\item[Table 2.] Comparison of our predictions for the sum rules
with the experimental values and with the quark parton model (QPM)
predictions without QCD corrections.
\end{itemize}

\bigskip

{{\Large \bf Figure captions}}

\bigskip

\begin{itemize}
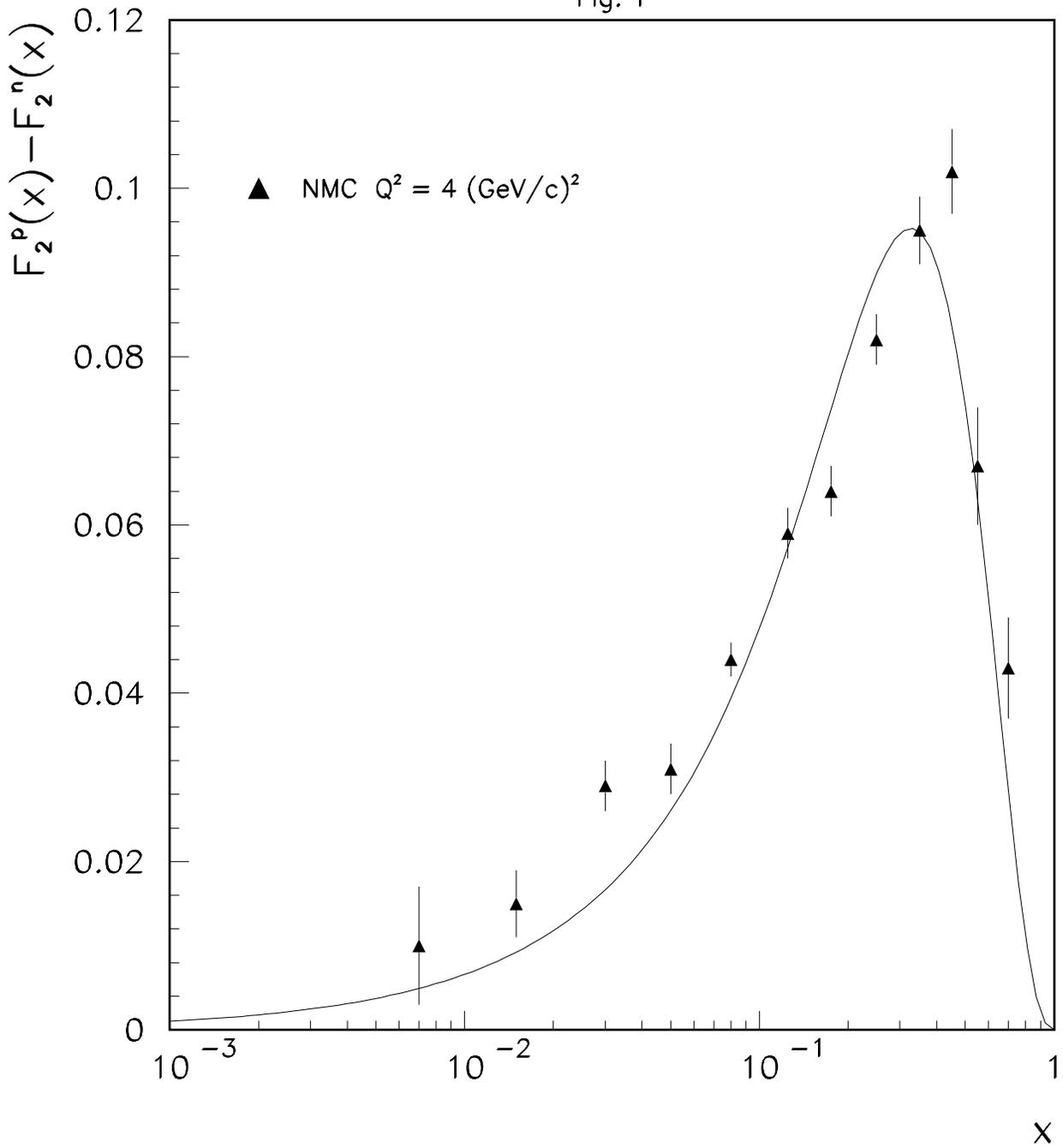
\item[Figure 1.] The prediction for $\fpx-\fnx$ is plotted and compared with
the experimental data \cite{7}.
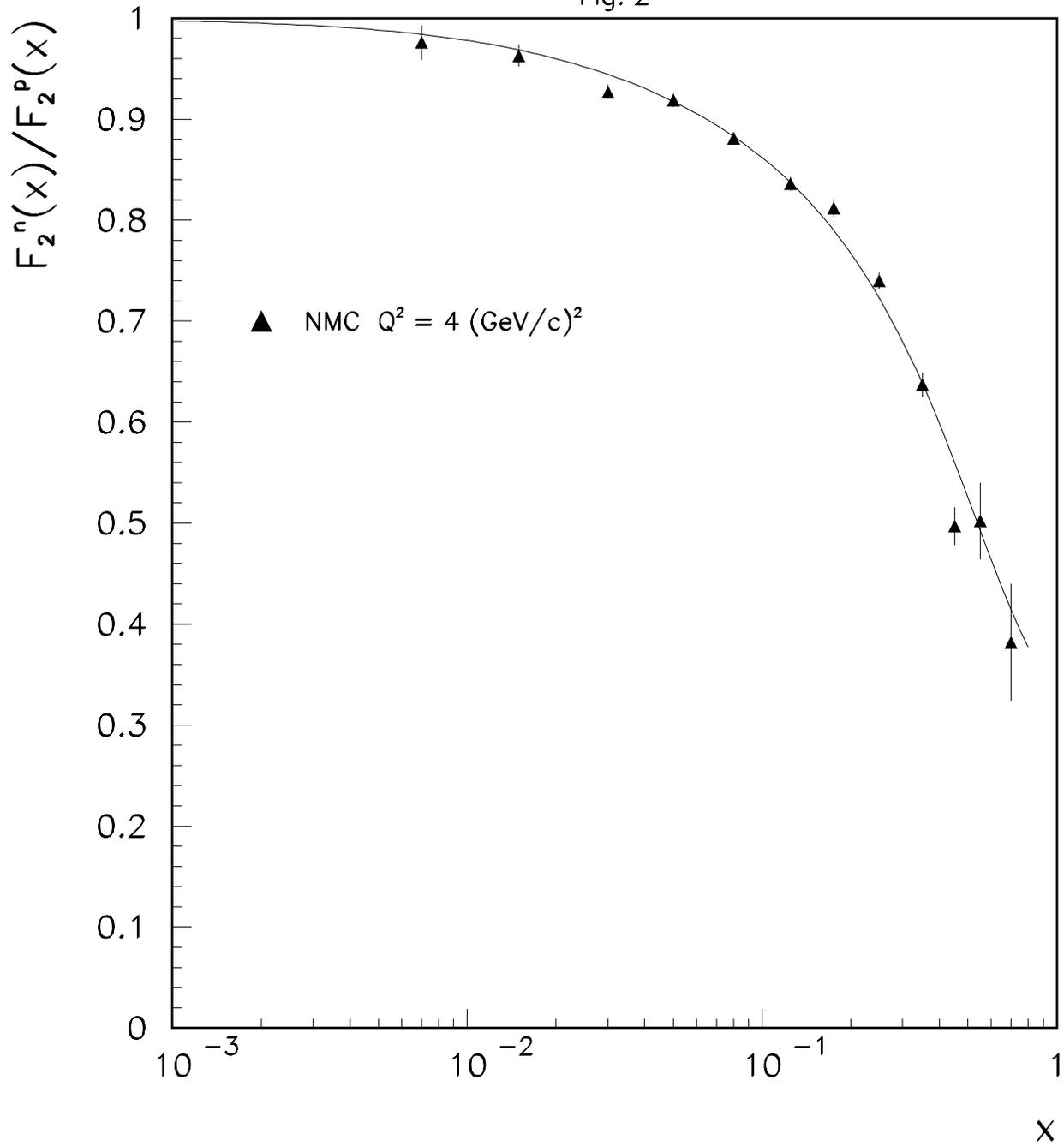
\item[Figure 2.] The prediction for $\fnx/\fpx$ is plotted and compared with
the experimental data \cite{7}.
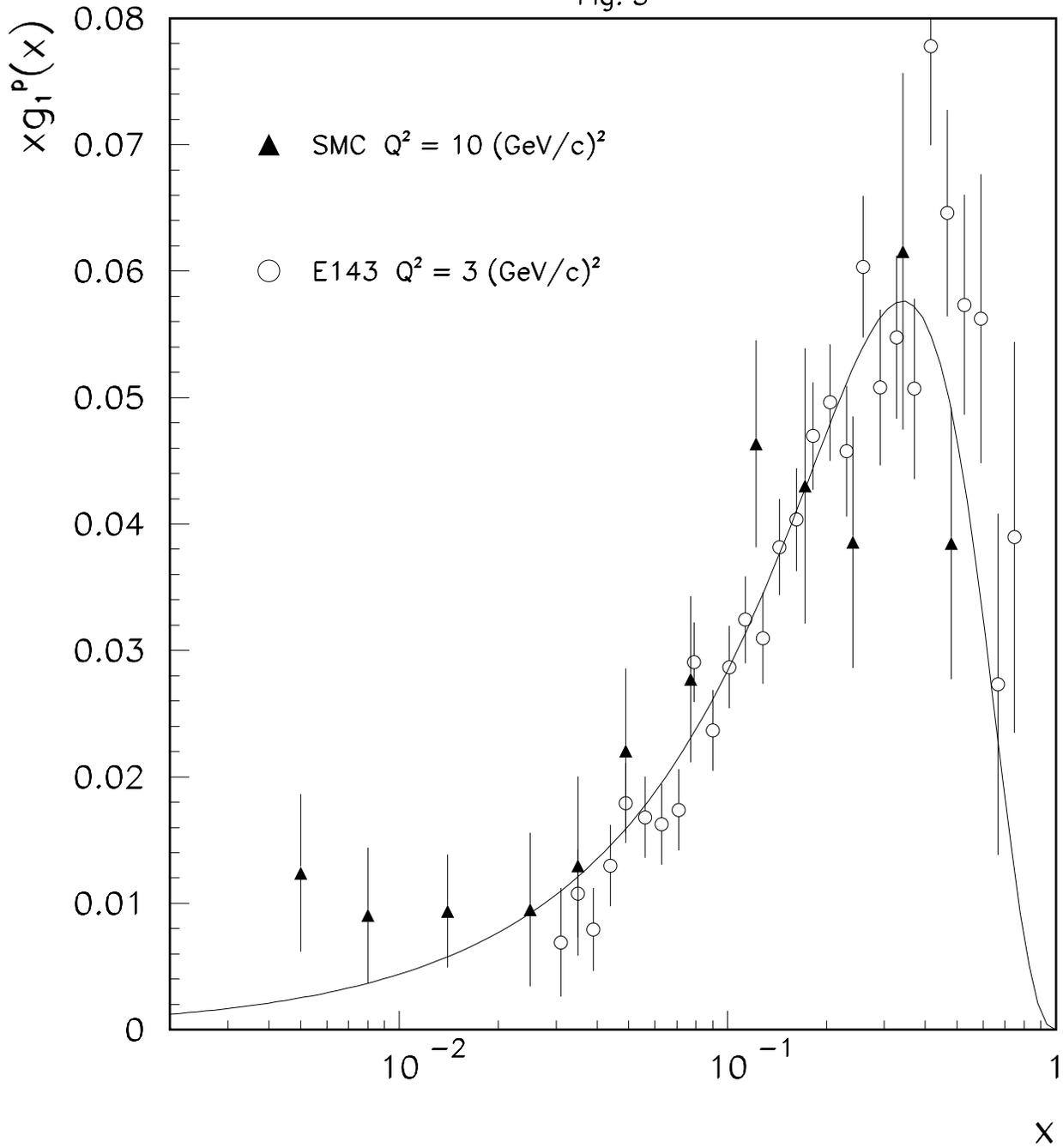
\item[Figure 3.] $x\gpx$ is plotted and compared with the data \cite{2},
\cite{2bis}.
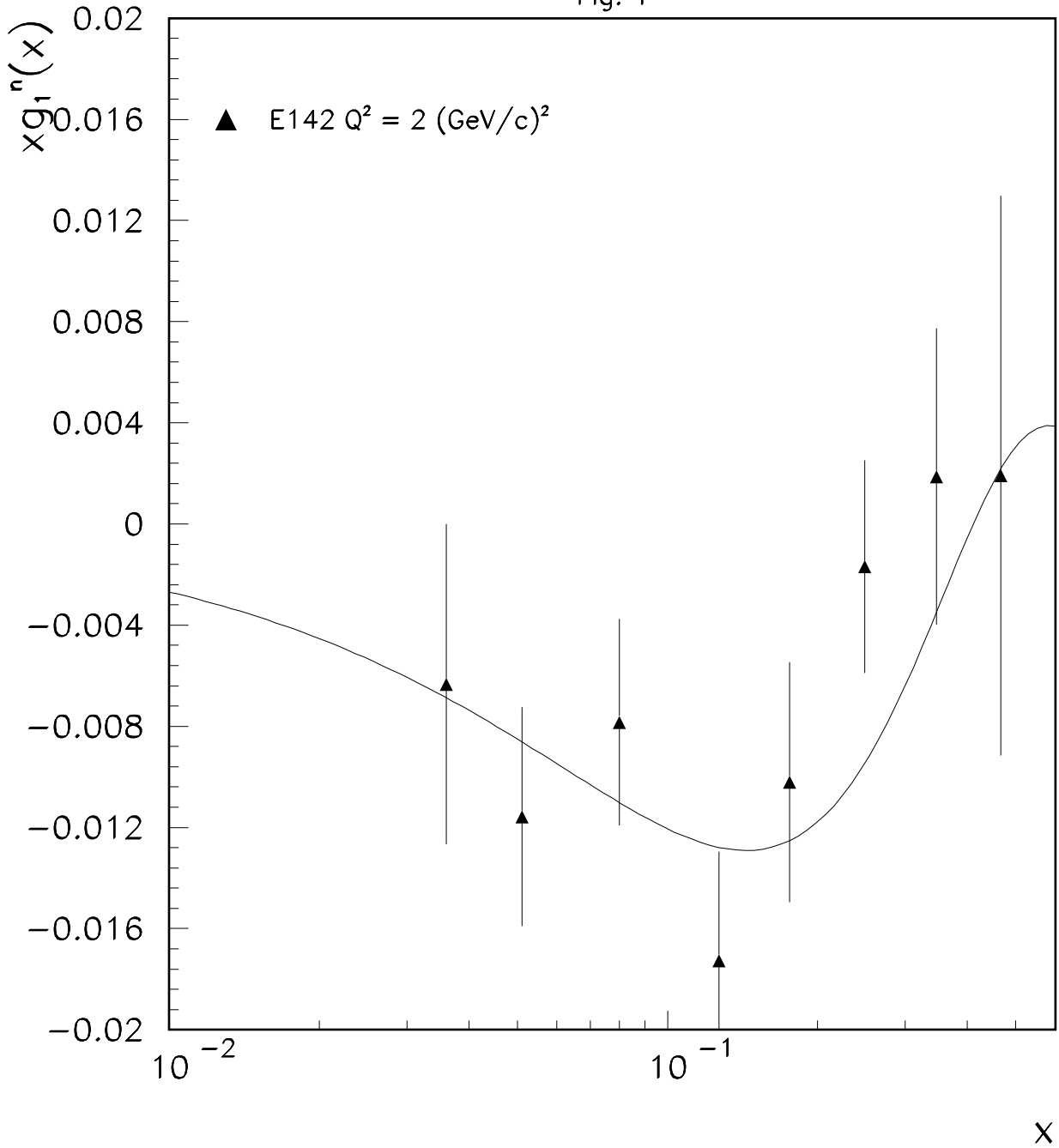
\item[Figure 4.] $x\gnx$ is plotted and compared with the data \cite{16}.
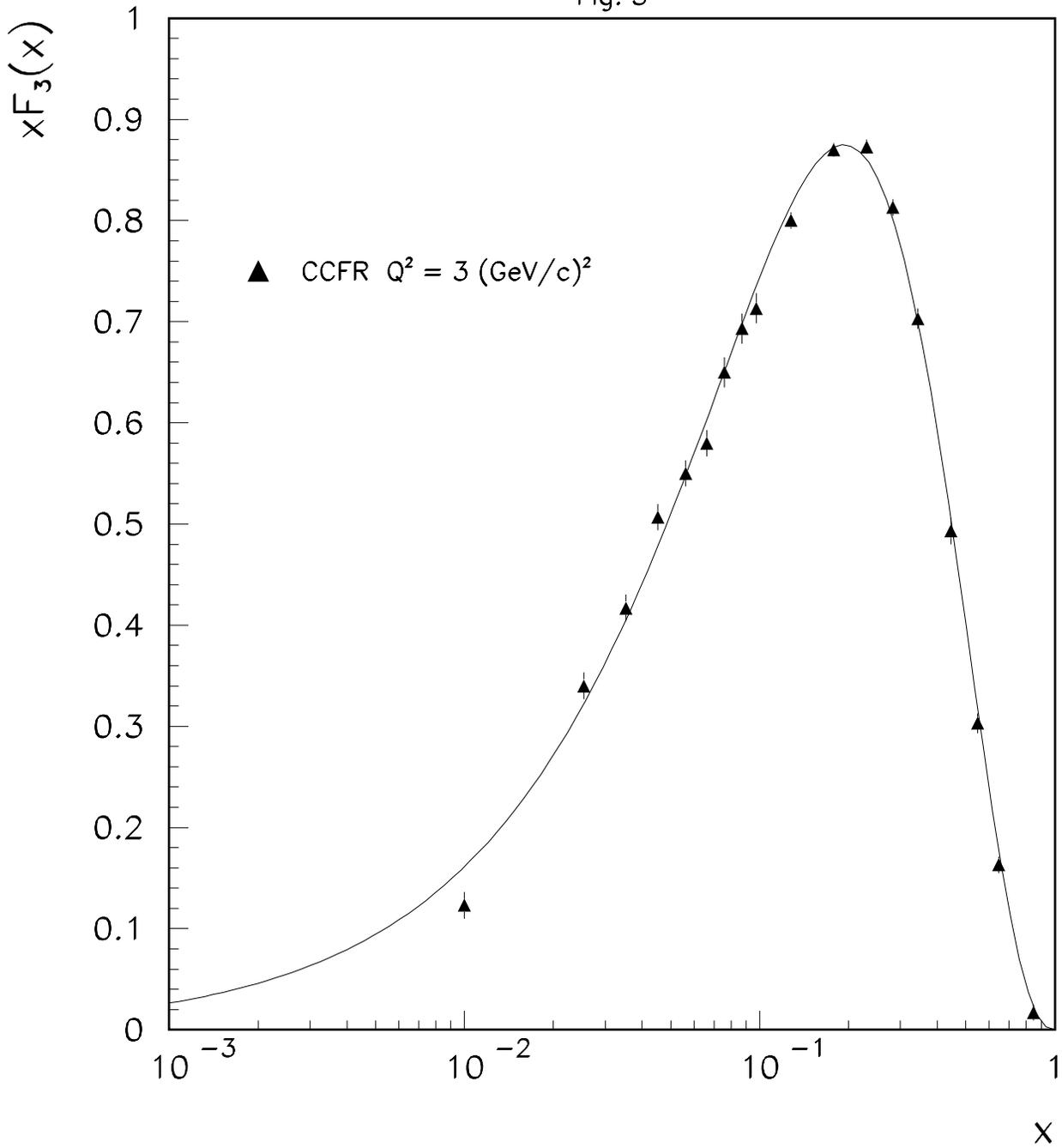
\item[Figure 5.] $xF_{3}(x)$ is plotted and the experimental values are taken
from \cite{12}.
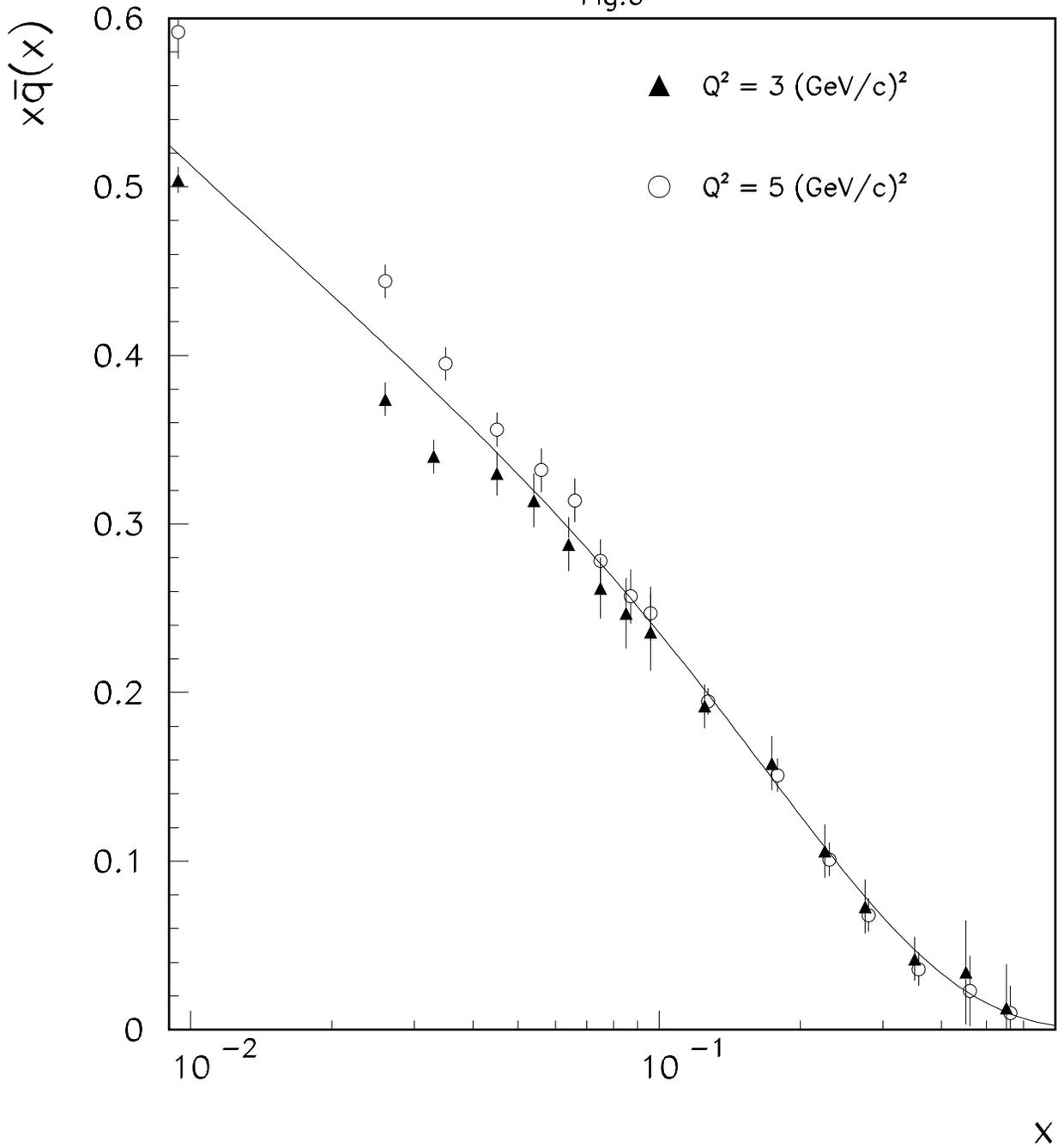
\item[Figure 6.] $x\bar{q}(x)$ versus $x$ is shown, the experimental data
correspond to \cite{9}.
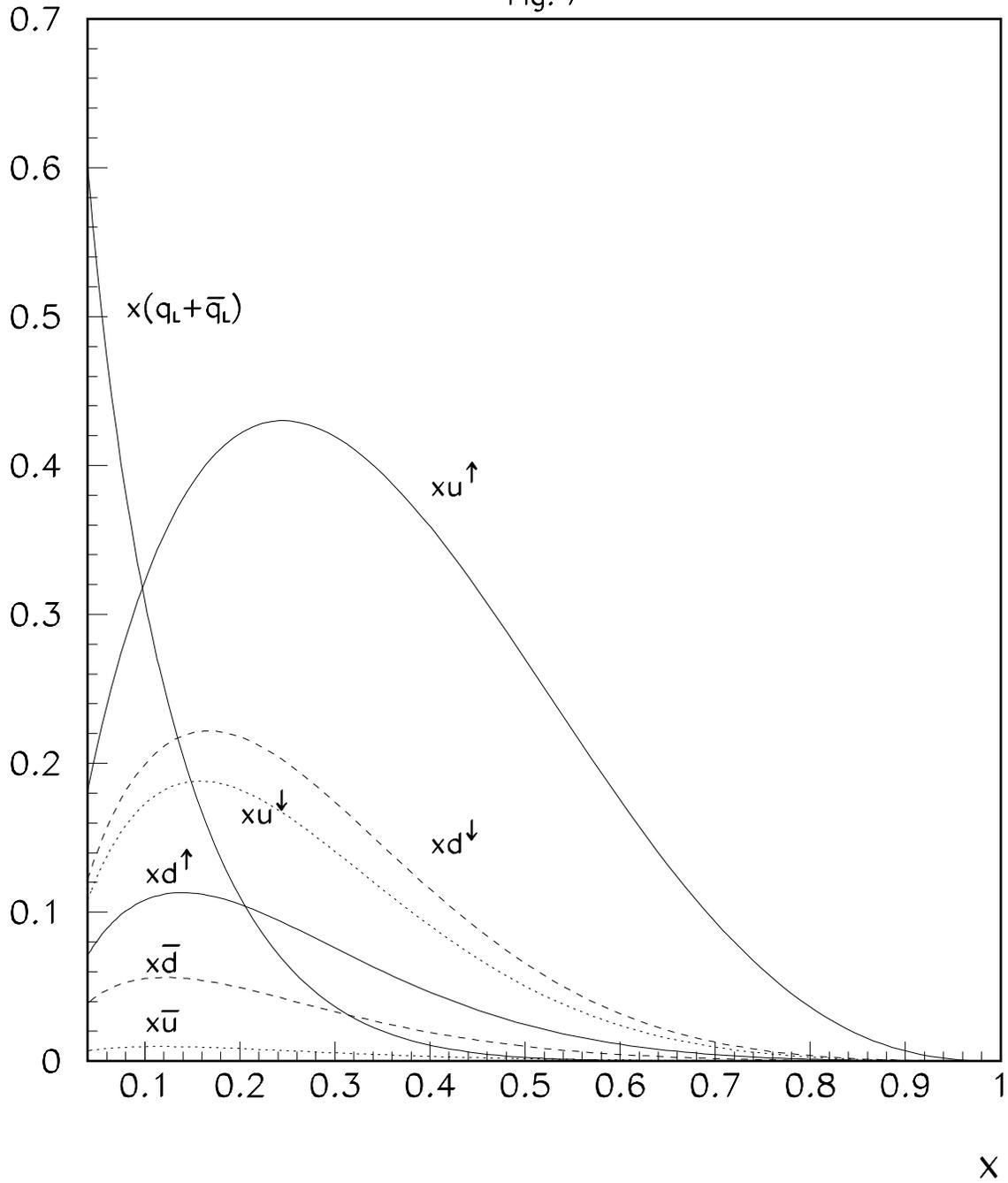
\item[Figure 7.] The momentum distributions of {\it gas} component of
$q$ and $\bar{q}$'s, and of the total {\it liquid} part are here shown.
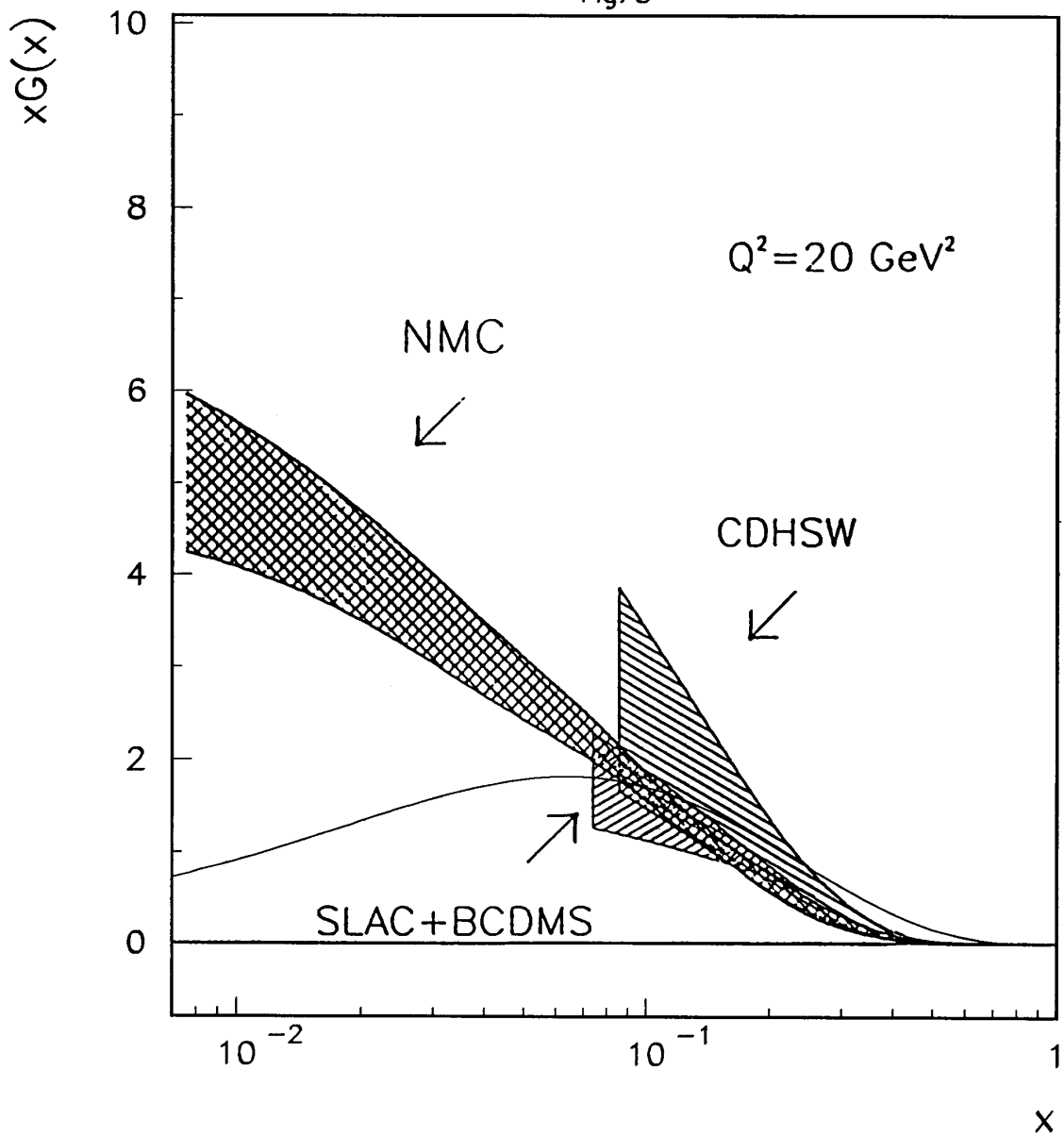
\item[Figure 8.] $xG(x)$ versus $x$ is shown, the experimental data
correspond to CDHSW \cite{CDHSW},\\ SLAC+BCDMS \cite{SLAC&BCDMS} and NMC
\cite{NMCgluon}.
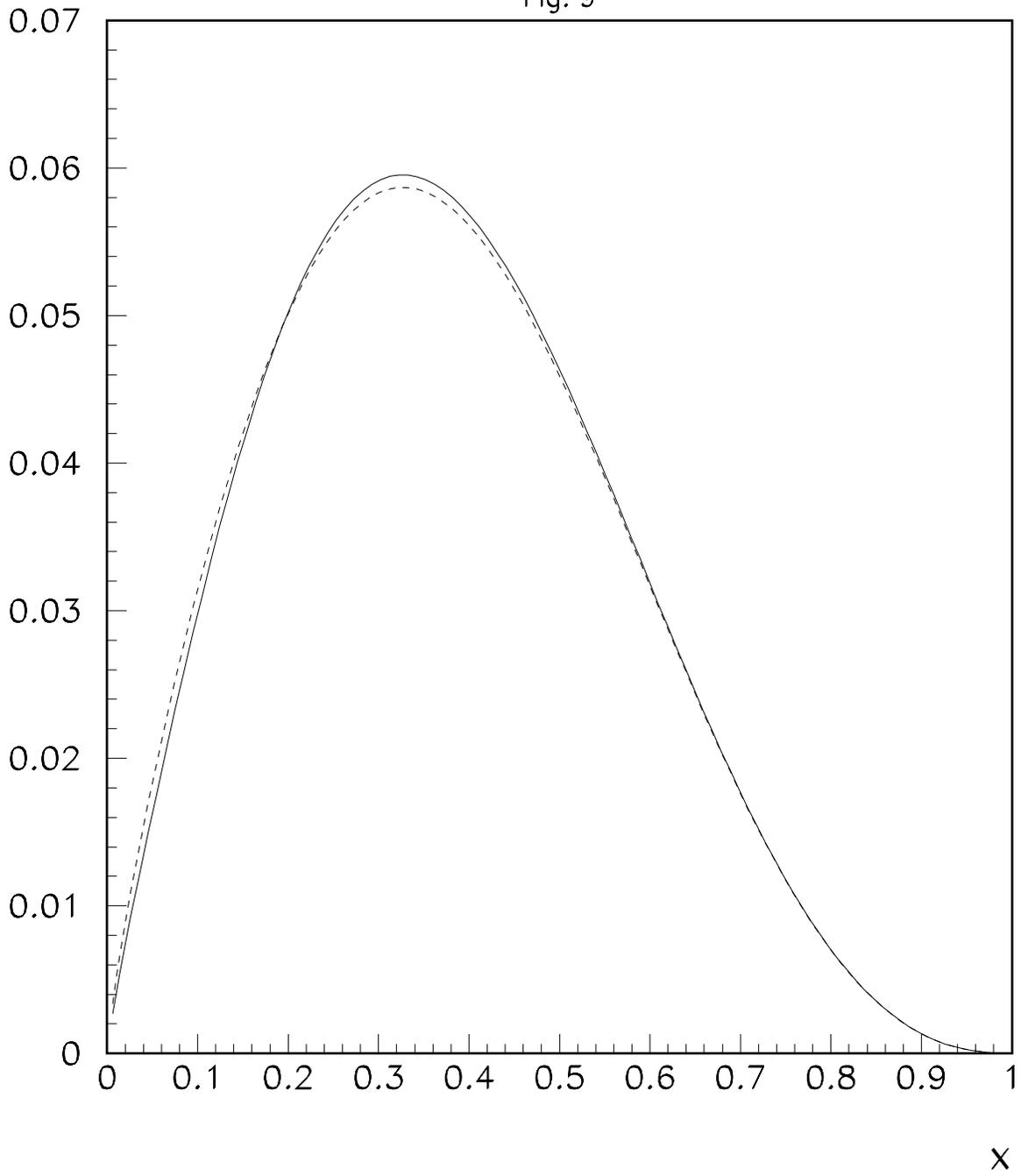
\item[Figure 9.] The predicted values for $x\left[ \gpx - { 1 \over 4}
\gnx \right]$ (dashed line), and ${ 5 \over 8} \left[\fpx - \fnx\right]$
(full line) are compared.
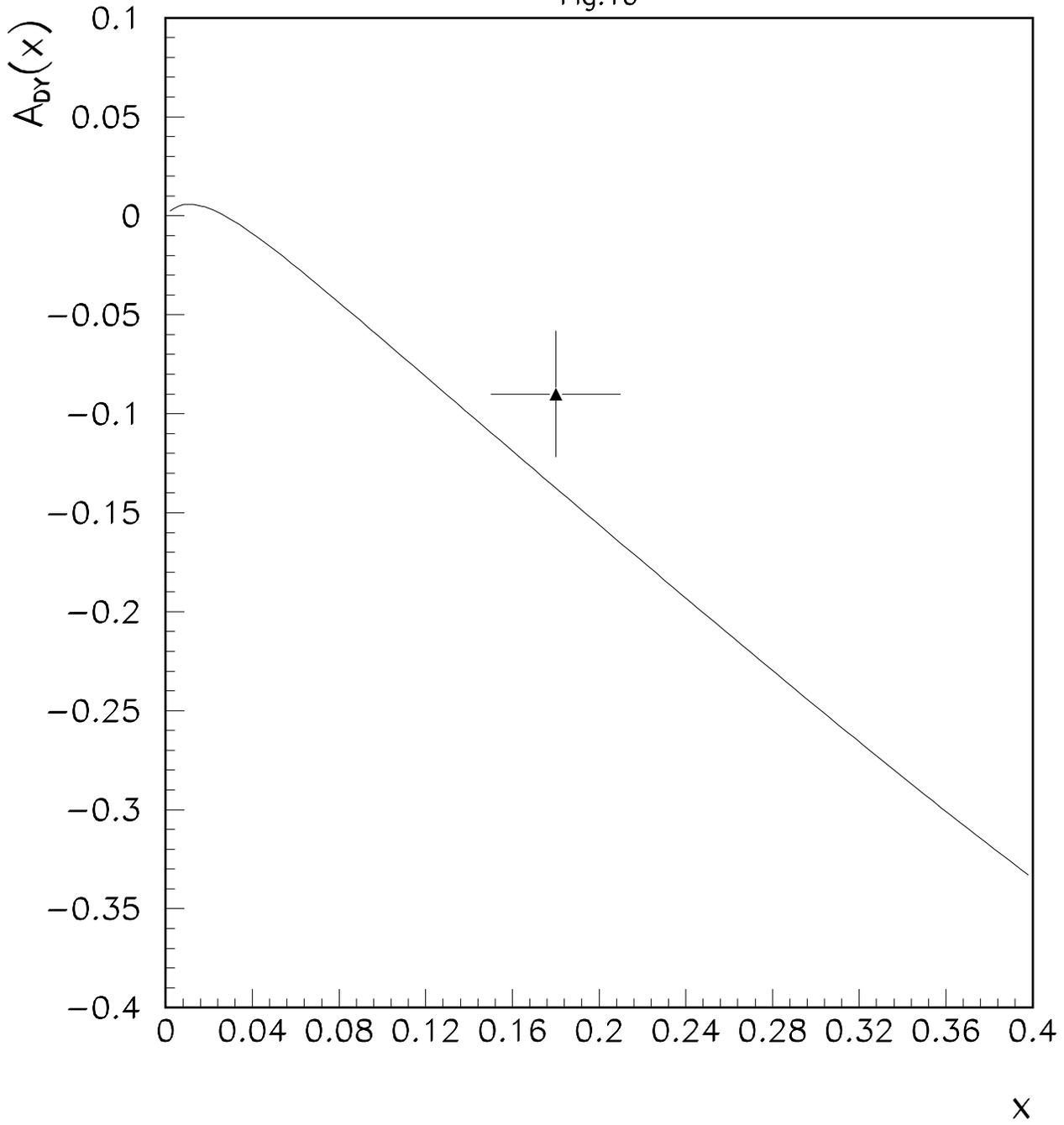
\item[Figure 10.] The asymmetry $A_{DY}(x)$ is here plotted,
the experimental result is taken from \cite{NA51}.
\end{itemize}
\end{document}